\documentstyle[aps,prb,multicol,epsfig]{revtex}

\begin{document}

\title{Scaling properties of one-dimensional Anderson models in an
electric field: Exponential vs. factorial localization}
\author{M.~Weiss, T.~Kottos, and T.~Geisel\\
Max-Planck-Institut f\"ur Str\"omungsforschung and
Fakult\"at f\"ur Physik der Universit\"at G\"ottingen,
Bunsenstr.~10, 37073~G\"ottingen, Germany}
\date{\today}
\maketitle

\begin{abstract}
We investigate the scaling properties of eigenstates of a one-dimensional
(1D) Anderson model in the presence of a constant electric field. The
states show a transition from exponential to factorial localization. For
infinite systems this transition can be described by a simple scaling
law based on a single parameter $\lambda_{\infty} = l_{\infty}/l_{\rm el}$,
the ratio between the Anderson localization length $l_{\infty}$ and 
the Stark localization length~$l_{\rm el}$.
For finite samples, however, the system
size $N$ enters the problem as a third parameter. In that case the global
structure of eigenstates is uniquely determined by two scaling
parameters $\lambda_N=l_\infty/N$ and $\lambda_\infty=l_\infty/l_{\rm el}$.
\end{abstract}


\begin{multicols}{2}
%------------------------------------------------------

\section{\bf Introduction}

In recent years several studies have investigated the influence of
constant uniform electric fields on the localization of electrons in
one-dimensional (1D) systems with on-site randomness. In the absence of
dc-fields it is by now well known that even small amounts of disorder lead
to an exponential localization of all eigenstates~\cite{A58,book}. On the
other hand, in the case of a single-orbital tight-binding model of an
electron in a periodic potential, application of a static electric field
is known to generate a discrete, uniformly spaced eigenvalue spectrum~\cite{W60}, 
known as Stark ladder, with all eigenfunction 
localized factorially~\cite{LL86}. For weak fields the wavefunctions may be extended
over several lattice periods, but with increasing field strength the
electron tends to be localized on a specific site. This is known as Stark
localization and has been observed experimentally in superlattices
\cite{MRH88RR} which are commonly used for such measurements~\cite{Y94}.

In infinite samples the localization of eigenstates can be characterized
in terms of the localization length; the latter is commonly defined from
the amplitude decay of eigenstates in the limit $|n|\rightarrow \infty$,
where~$n$ labels the site in the tight-binding picture. The most powerful
and informative method available for such studies is the transfer matrix 
method. In the presence of a strong electric field, however, it appears
to be less efficient due to the factorial nature of the Stark localization. 
Moreover, for finite systems the structure of eigenvectors cannot be
characterized in the same way. One then needs to use other quantities (such as
the inverse participation ratio), that are valid both for finite and infinite
samples. Through the use of scaling conjectures, one can link then the
properties of eigenstates in infinite samples to those of finite samples.
Since the scaling approach proved to be extremely useful in describing
conductance and its fluctuations (see, e.g.,~\cite{AALR79,P86}) in the
theory of disordered solids, it seems natural to use this approach also in
order to describe localization properties of eigenfunctions of 1D disordered
systems in the presence of constant electric field.

In this paper we study the 1D Anderson model in the presence of a constant
electric field in view of scaling properties of its eigenstates. The main 
question that we want to answer is whether the up to now known equivalence 
between quasi-1D and 1D disordered models \cite{K94,FM94} can be extended 
in order to include also systems with constant electric field. For this 
purpose we analyze the scaling properties of information lengths for infinite 
and finite samples, which were used successfully in the studies of one and 
quasi-one dimensional systems~\cite{FM91,CGIFM92,M93,IKT96,FM94,CCGI93,FGIM93}. 
Contrary to the standard Anderson case, where the ratio of the Anderson 
localization length~$l_{\infty}$ and the sample size~$N$, i.e.~$\lambda_N= 
l_\infty/N$, is the only relevant scaling parameter, we find in the present 
case an additional scaling parameter~$\lambda_\infty= l_\infty/l_{\rm el}$. 
Here $l_{\rm el}$ is the Stark localization length, which arises from the 
applied constant electric field. Hence, the structure of eigenvectors in 
our model is characterized by two scaling parameters $\lambda_N,\,
\lambda_\infty$. 

The structure of the paper is as follows. In Section~2 we describe
the mathematical model and briefly summarize the known facts about the
two limiting cases that appear for our model. In Section~3 we discuss
different definitions of localization length, which are used in our numerical
simulations. In Section~4 we present numerical data on scaling of localization lengths
of eigenstates in infinite and finite systems. 
Finally, in Section~5 we study the scaling of the whole distribution of eigenvectors.
Our conclusions are summarized in Section 6.

%----------------------------------------------------------------------------------
\section{\bf The Model}
%----------------------------------------------------------------------------------

Our starting point is the 1D Schr\"odinger equation in the tight-binding
approximation
\begin{equation}
\label{tbe}
i{\frac{d\psi_n(t)}{dt}}=\,(\epsilon _n+neF) \psi_n(t)+V 
\psi_{n+1}(t)+ V\psi_{n-1}(t) \,,
\end{equation}
where~$\psi_n(t)$ denotes the probability amplitude for an electron to be at
site~$n$. Moreover, $\epsilon_n$ is the local site energy, $V$~is the hopping
element, $e$ is the electron charge and $F$ the strength
of the applied dc field. By applying the transformation $\psi_n(t)=\exp
(-i\,E\,t)\,\varphi_n$ one obtains the stationary equation
\begin{equation}
\label{tbe2}E\varphi _n=V \varphi _{n+1}+(\epsilon _n+neF)\varphi _n+
V \varphi_{n-1}\,,
\end{equation}
for the eigenvalues $E$ and the corresponding eigenstates~$\varphi _n(E)$.
We can distinguish two limiting cases which are relevant for our study: (a)
perfect system (i.e. $\epsilon_n = \epsilon$) with non-zero electric field
$F\neq 0$, and (b) zero field (i.e. $F=0$) with random on-site potential.

\begin{figure}
\begin{center}
\epsfig{figure=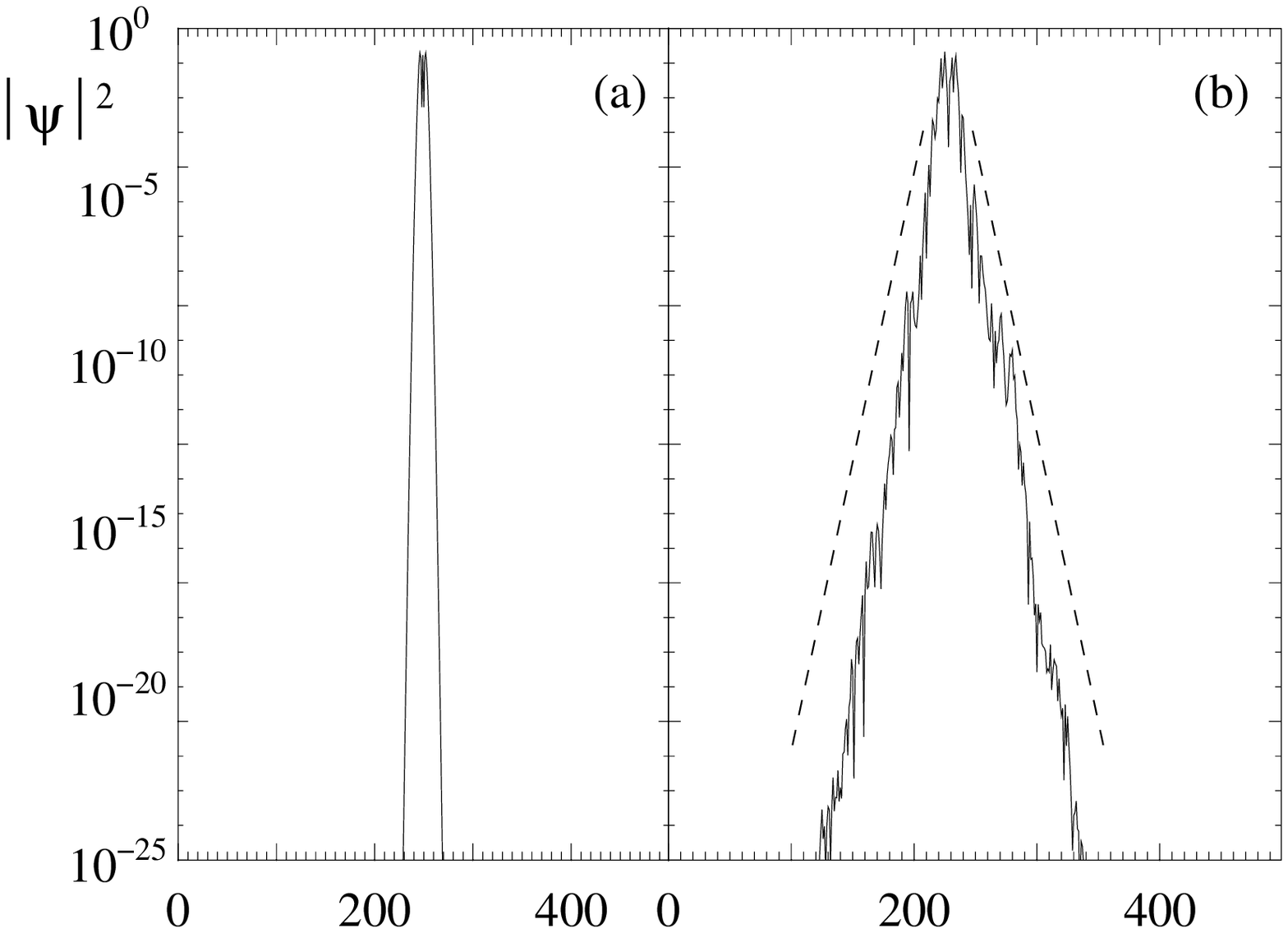,width=8cm,angle=360}\hfill\\
\end{center}
\noindent
{\footnotesize {\bf FIG. 1.} Two representative eigenfunctions of the 1D tight-
binding model (\ref{tbe}). {\bf (a)} Stark regime with factorial localization 
($W=0,\, eF=0.5,\,l_{\rm el}\approx 2$). {\bf (b)} Anderson regime with exponential 
localization ($W=5,\,eF=0,\,l_\infty\approx 3.4$). The dashed line has slope
$2/l_{\infty}$.
}
\end{figure}

In the case of a perfect system with $F\neq 0$, one can prove that the
corresponding
eigenstates $\varphi(E)$, known as Wannier-Stark states, show a generic
factorial decay i.e. \cite{LL86,RM82}
\begin{equation}
\label{wse}
\varphi_n(E) =J_{m-n}(2/eF) \rightarrow (1/eF)^{|n|}/(|n|!) ; n\rightarrow
\pm\infty
\end{equation}
where $J_n$ is the Bessel function of order $n$. Wannier-Stark states
constitute a complete set of energy eigenstates~\cite{FBF73}. Their
eigenvalues,
$E_m = m e F$ form the so-called Wannier-Stark ladder~\cite{W60}. A particular
Wannier-Stark state $\varphi_n$ is factorially localized around the $n-$th
site, with a localization length of the order of $l_{\rm el}\simeq 1/eF$, i.e.
the electric field appears in the denominator of the localization length
in Eq.~(\ref{wse}). This underscores the fact that $F$ cannot be treated as
a small perturbation to the field-free Hamiltonian. An example of such a state
is presented in Fig.~1a. 

The other limit of interest corresponds to zero electric field with $\epsilon_n$
random and $\delta$-correlated, chosen from a distribution ${\cal P}_{\epsilon}$
with mean zero and variance $\sigma^2$. Below, in our numerical investigation
we will always assume that ${\cal P}_{\epsilon}$ is a uniform distribution
in $[-W/2,W/2]$.  Such a model is known in the literature
as the Anderson model~\cite{A58} and has been studied in great detail in the
context of disordered materials. It was shown with mathematical rigor that
in the limit of infinite samples this model displays exponentially localized
eigenfunctions, no matter how small the disorder is (see Fig.~1b). The rate
of decay is measured by the Lyapunov exponent~$\gamma$ which may be evaluated
by the transfer matrix method. To this end, one has to study the asymptotic
behavior of the random matrix product~$\prod {\em M}_n$, where~{\em M}$_n$ is
defined through the relation
\begin{equation}
\label{M}\xi _{n+1}={\em M}_n\xi _n;\,\,\,\,\,\,\,{\em M}_n=\left(
\begin{array}{cc}
v_n & -1 \\
1 & 0
\end{array}
\right) ;\,\,\,v_n=\frac{E-\epsilon _n}{V}\,\,
\end{equation}
for the vector~$\xi _n=(x_n,x_{n-1})\,$ with the matrix~${\em M}_n$ known
as the transfer matrix. The localization length $l_\infty $ is hence the
inverse Lyapunov exponent~$\gamma $; the latter is evaluated as the exponential
rate of increase of an initial vector~$\xi_1$,
\begin{equation}
\label{gamma}l_\infty ^{-1}=\gamma =\lim _{N\rightarrow \infty }\frac 1N
\frac{\ln \left(\,\prod\limits_{n=1}^N|{\em M}_n\xi _n|\right )}{|\xi _1|}\,.
\end{equation}
Although the Lyapunov exponent~$\gamma $ for finite~$N$ depends on the
particular
realization of disorder, for $N\rightarrow \infty$ it converges to its mean
value \cite{O68}. For the calculations below we have used samples of length
$5\times 10^5$ for relatively large values of $W$ and up to $4\times 10^6$
for small values of $W$ .

%----------------------------------------------------------------------------------
\section{\bf Scaling approach for the eigenstates}
%----------------------------------------------------------------------------------

Our interest is dedicated to the structure of eigenstates for infinite as 
well as finite samples, as we tune the disorder and the electric 
field strength. Unlike the simpler case of infinite samples, however, the
meaning of a localization length for finite samples is not clear. Below we
follow the approach developed in the theory of quasi-1D disordered solids
which is based on the evaluation of multifractal localization lengths (see,
e.g.,~\cite{FM94}). The great advantage of this approach is the applicability
in both finite and infinite samples.

One of the commonly used quantities in this approach is the so-called entropic
localization length, defined through the information entropy ${\cal H}_N$ of
eigenstates,
\begin{equation}
\label{HN}{\cal H}_N=-\sum\limits_{n=1}^Nw_n\ln
\,w_n;\,\,\,\,\,\,\,w_n=|\varphi _n^2|
\end{equation}
where $\varphi _n$ is the $n-$th component of an eigenstate in a given basis.
For eigenstates normalized as $\sum_nw_n=1$, the simplest case of $\varphi
_n=N^{-1/2}$ results in an entropy equal to the maximum value: ${\cal H}_N=
\ln (N)\,$. We therefore define the localization length~$L$ as the number of 
basis states occupied by the eigenstate $\varphi_n\,$; the latter is equal to 
$\exp ({\cal H}_N)$. We note that in general, the amplitudes $\varphi _n$ 
fluctuate strongly with~$n$ and thus the coefficient of proportionality between 
$L$ and $l_\infty \,$ depends on the type of fluctuations.

In order to study the properties of eigenstates in quasi-1D solids,
localized on some scale in the finite basis, it was suggested in~\cite{CGIS90}
to normalize the localization length~$L$ in such a way that in the extreme
case of fully extended states the quantity~$L$ is equal to the size of the
basis $N$. In such an approach, the entropic localization length $L_1$ is
defined as
\begin{equation}
\label{l1}L_1=N\exp (<{\cal H}_N>\,-\,{\cal H}_{\rm ref}\,)
\end{equation}
In Eq.~(\ref{l1}) the ensemble average $<...>$ is performed over the number
of eigenstates with the same structure and over realizations of the disorder
potential. The normalization factor~${\cal H}_{\rm ref}$ has the meaning of
an average entropy of the completely extended random eigenstates in the finite
basis. For the quasi-1D case the distribution of components~$\varphi_n$ is
assumed to correspond to the Gaussian Orthogonal Ensemble (GOE) ~\cite{CGIS90}.

Analogously, a whole set of localization lengths $L_q\,$ can be defined
in the following way~\cite{FM94}:
\begin{equation}
\label{lq}L_q=\,N\,(\frac{<P_q>}{P_{\rm ref}^{(q)}})^{\frac
1{1-q}}\,\,; \,\, P_q=\sum\limits_{n=1}^N(w_n)^q\,\,\,\,;
q\geq 2
\end{equation}
where \thinspace $P_{\rm ref}^{(q)}$ \thinspace is the average value of $P_q$
for the reference ensemble of completely extended states. One should note
that for the particular case~$q=2$ the quantity~$P_2$ is known as the
participation ratio.

In the context of quasi-1D disordered models in the presence of constant
electric field~\cite{FM91,FGIM93,CCGI93}, it was shown that all global properties
of eigenfunctions are described by the following localization parameters
\begin{equation}
\label{betaq}\,\beta_q^{\infty}=\frac{L_q}
{l_{\rm el}};\,\,\,\beta_q^{N}=\frac{L_q} {N}
\end{equation}
where the superscript $\infty\,(N)$ denotes infinite (finite) samples,
respectively. Moreover it was found that~$\beta_q^{\infty,N}$ obey some
scaling law, i.e. they depend only on the ratio of the characteristic lengths
of the system. In the case of infinite samples only two length scales, i.e. 
$l_{\infty}$ and $l_{\rm el}$, are relevant. For finite samples, however, a third
length~$N$, which is the actual size of the sample, comes into play and has to
be taken into account in the scaling theory. According to the scaling conjecture 
in the modern theory of disordered solids, it was found that for quasi-1D
systems in the presence of electric field \cite{FM91,FGIM93,CCGI93} the
$\beta_q^{\infty,N}$
follow the scaling laws:
\begin{eqnarray}
\label{scl} \beta_q^{\infty} =
\beta_q^{\infty}(\lambda_{\infty})&;&\,\,\,\,
\beta_q^{N}=\beta_q^{N}(\lambda_{\infty},\lambda_N),\,\,\, \nonumber \\
{\rm where} \quad\lambda_{\infty} =\frac{\l_{\infty}}{l_{\rm
el}}&;&\,\,\,\,
\lambda_{N} =\frac{\l_{\infty}}{N}.
\end{eqnarray}

Our main question is whether relations of the type of Eq.~(\ref{scl}) are 
also applicable for our 1D Anderson tight-binding model with electric field. 
In Refs.~\cite{FM94,K94} is was shown analytically that the eigenstates in 
1D and quasi-1D disordered systems {\it without} electric field, possess the 
same gross structure (envelope) on scales comparable with the localization 
length, while their statistical properties are quite different on a much finer 
scale of the order of the lattice spacing. That is the reason why many scaling 
laws, which are dominated by the fluctuations of the envelope, hold both for 1D 
and quasi-1D.
The validity of such a statement is however questionable in the presence
of electric fiels. We will show that such a similarity between quasi-1D and
1D disordered systems persist also in this case.

The first nontrivial question in this context arises about the reference 
ensemble for the computation of the average entropy ${\cal H}_{\rm ref}$. 
Indeed, in application to 1D Anderson type models \cite{CGIFM92,M93,IKT96}
the reference ensemble cannot be chosen as an ensemble of full random matrices, 
like the GOE.  This point is related to the fact that in 
the 1D tight-binding case fully extended states are not Gaussian random 
functions but just plane waves which arise for zero disorder. In the presence 
of electric field, the situation is even more complicated due to strong 
dependence of the eigenstates on the electric field. However, and this is 
our expectation, in spite of the above mentioned differences, scaling 
properties of the eigenstates of the 1D model (\ref{tbe}) are of the generic 
type discovered for quasi-1D systems.

For this reason and in the spirit of Refs.~\cite{CGIFM92,IKT96}, we define the
normalization factors ${\cal H}_{\rm ref}$ and $P_{\rm ref}^{(q)}$ from the
solution of Eq.~(\ref{tbe2}) for zero disorder and electric field, i.e.
$\epsilon_n=0$ and $F=0$
%\begin{eqnarray*}
\begin{equation}
\label{EF}
%\label{eigen}
E^k=2V\cos \frac{k\pi }{N+1}\,,\nonumber
%\end{eqnarray*}
%\begin{equation}
\varphi _n^k=\sqrt{\left( \frac 2{N+1}\right) }\sin \frac{nk\pi
}{N+1}\,,
\end{equation}
with $k,n=1,\dots ,N$. The entropy ${\cal H}_{\rm ref}$ and the participation
ration $P_{\rm ref}^{(2)}$ of the above eigenstates in the large~$N$ limit 
has the same value for every eigenvalue~$E^k$, i.e.
\begin{equation}
\label{norm}{\cal H}_{\rm ref}=\ln (2N)-1\,\,;\,\,P_{\rm ref}^{(2)}=\frac 3{2N}.
\end{equation}

%------------------------------------------------------------------------------
\section{\bf Scaling properties of localization lengths}
%------------------------------------------------------------------------------
\subsection{\bf Infinite samples}

In this Section we analyze the scaling properties of eigenstates of infinite
systems.  In numerical studies the matrices are obviously of finite size $N$.
However in our analysis below we will always consider the case that $N\gg l_{\infty},
\, l_{\rm el}$, and thus the finite (but large) size of the matrix becomes
irrelevant. We therefore have used these data to investigate our
scaling assumption for $\beta_q^{\infty}$ (see Eq.~(\ref{betaq})).

As was mentioned in Section~II the introduction of a non-zero electric field
$F\neq0$, results in an additional length scale $l_{\rm el}$. This length
arises
when we consider a cross section of the energy band locally tilted by the
electric
field: $-V/2+ F n\leq E \leq V/2 + Fn$ for an energy level $E.$ Therefore
the scaling parameter $\lambda_{\infty} =l_{\infty}/l_{\rm el}$ enters the problem.
Furthermore, if we consider the localization lengths $L^{(q)}$ of
Eqs.~(\ref{l1}),(\ref{lq}) as the typical length, which contains most of an eigenvector
normalization, we expect that
\begin{equation}
\label{xi1}
L_q \simeq
\begin{array}{cc}
l_{\infty}& \lambda_{\infty}\ll 1\nonumber \\
l_{el}    & \lambda_{\infty}\gg 1
\end{array}\,\, ,
\end{equation}
i.e. the exponentially localized states progressively becomes
localized factorially as the field increases. This is due to the fact that, 
for weak electric field we have $\lambda_{\infty}\ll 1$, and thus the dominant 
localization mechanism, i.e. the one that produces the shortest localization
length scale, is the one related to the randomness. From now on we will refer 
to this as the "Anderson regime".  In the opposite limit $\lambda_{\infty}\gg 1$, 
the dominant localization mechanism is due to the electric field. We will refer 
to this regime as the "Stark regime". Based on the previous considerations we 
expect that the $L_q$'s have the following scaling form, (see also Ref.~\cite{FM91} 
for an equivalent reasoning for quasi-1D systems) 
\begin{equation}
\label{xi2}
L_q = l_{\infty} f(\lambda_{\infty})\,\,\,{\rm with}\,\,\, f(\lambda_{\infty}) \simeq
\begin{array}{cc}
 ~ 1                      & \lambda_{\infty} \ll 1 \nonumber\\
\frac 1{\lambda_{\infty}} & \lambda_{\infty} \gg 1
\end{array}
\end{equation}
where $f(\lambda_{\infty})$ is related to the scaling function $\beta_q^{\infty}$
as $\beta_q^{\infty}= \lambda_{\infty} f(\lambda_{\infty})$.
\begin{figure}
\begin{center}
\epsfig{figure=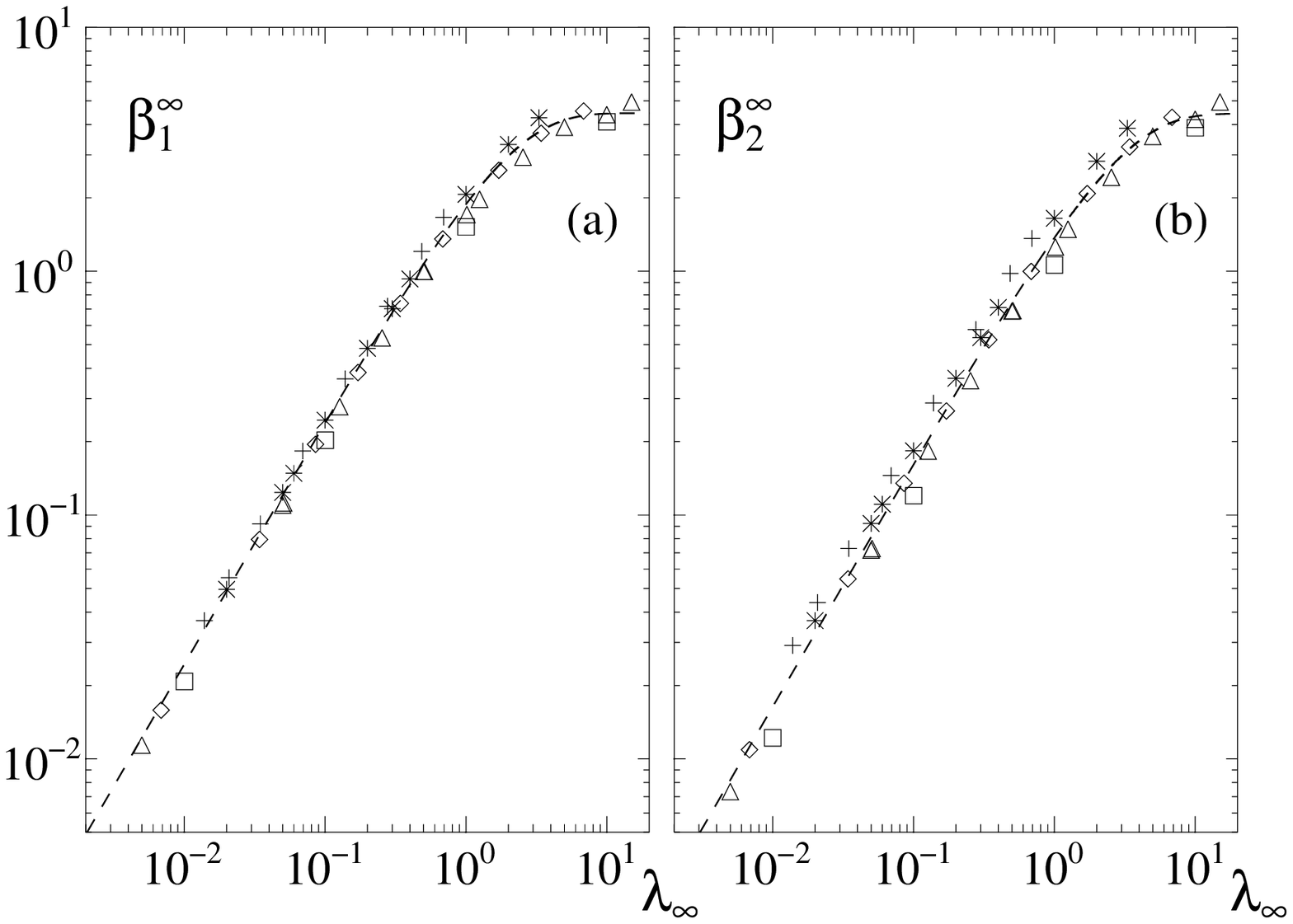,width=8cm,angle=360}\hfill\\
\end{center}
\noindent
{\footnotesize {\bf FIG. 2.} One-parameter scaling of the (nearly) infinite sample 
($N=10^4$) upon variation of~$\lambda_\infty$ in a range~$N\gg l_{\rm el},l_\infty$
($W=2.62,3.87,5,7.35,10, eF\in[10^{-4},2]$). 
A least squares fit according to Eq.~(\ref{fit1}) is shown as dashed line.
{\bf (a)} Scaling of~$\beta^\infty_1$ ($a_1^0=4.45$ and $a_1^1=0.55$).
{\bf (b)} Scaling of~$\beta^\infty_2$ ($a_2^0=4.43$ and $a_2^1=0.37$).
}
\end{figure}

Our aim in this paragraph is to support the above mentioned scaling law 
based on numerical data and to extend our knowledge on the structure of 
the eigenstates in the
intermediate regime between the two discussed limits. In order to study scaling
properties of the localized eigenstates we have used the transfer matrix method 
for the calculation of $l_{\infty}$ as well as the direct diagonalization of the 
Hamiltonians that are associated with Eq.~(\ref{tbe}), for finite but long chains 
of size~$N=10^4$. In all numerical calculations below we used~$V=1$.
We then varied the disorder strength $W$ as well as the dc field 
strength in a regime, where always $l_{\infty},l_{\rm el}\ll N$. One should 
stress here that both localization lengths $l_{\infty} $ and $L_q$ are functions 
of the energy $E$. For this reason, in our numerical experiments we consider 
ensembles of states specified by the values of the energy $E$ in a small window 
$E\in[0.95,1.05]$. The size of the energy window was chosen in such a way that the 
localization length $l_\infty $ is approximately constant inside this window (in 
all the cases $\frac{\Delta l_\infty } {l_\infty }\leq 0.06$ ). The values of 
$\beta^\infty_1$ and $\beta^\infty_2$ are then obtained by averaging over all 
eigenstates which are found inside the energy window for a set of different 
realizations of disorder. As a result, the total number of eigenstates used for 
the calculation of $\beta^{\infty}_q$ were more than $1500$.

A detailed analysis of the numerical data gives evidence for a scaling 
behaviour of the form of Eqs.~(\ref{scl}),(\ref{xi2}) with the scaling function
\begin{equation}
\label{fit1}
\beta_{q}^{\infty} \approx a_q^0 (1 - \exp(-a_q^1 \lambda_{\infty}))
\end{equation}
where the parameters $a_q^0,a_q^1$ are determined from a least squares fit.
We have found that $a_q^0 = 4.45~(4.43)$ and $a_q^1=0.55 ~(0.37)$
for $q=1~(2)$. We notice here, that a similar scaling function was found in the
framework of quasi-1D systems for $q=1$ \cite{CCGI93}. Our data together with a fit 
according to Eq.~(\ref{fit1}) are presented in Fig.~2. We observe a nice agreement 
with the scaling assumption of Eqs.~(\ref{scl}),(\ref{xi2}).

%------------------------------------------------------------------------------
\subsection{\bf Finite samples}
%------------------------------------------------------------------------------

In realistic situations one always deals with finite samples. In such cases
the understanding of the statistical properties of conductance are of major
importance. Since these properties are directly related to the structure
of eigenstates, it is important to investigate the statistical properties of
eigenstates for finite systems. This is the goal of the present subsection.

For finite $N$ and zero electric field, it was shown in \cite{CGIFM92,M93,IKT96},
that the statistical properties of 1D Anderson models are characterized by
a single scaling parameter $\lambda_N =l_{\infty}/N$. Moreover, the scaling
relation for the eigenvectors was found to be very simple
\begin{equation}
\label{fse}
\beta_q^N =\beta_q^N(\lambda_N)=\frac{c_q \lambda_N}{1+c_q\lambda_N}
\end{equation}
where the constants $c_q$ were found to be $c_1\approx 2.6$ and $c_2\approx 1.5$.
In fact, this scaling relation is exact only for $q=2$.
For other cases of small values~$q$, including $q=1$, however, it is very close to the
correct one (see details in~\cite{FM94}).
\begin{figure}
\begin{center}
\epsfig{figure=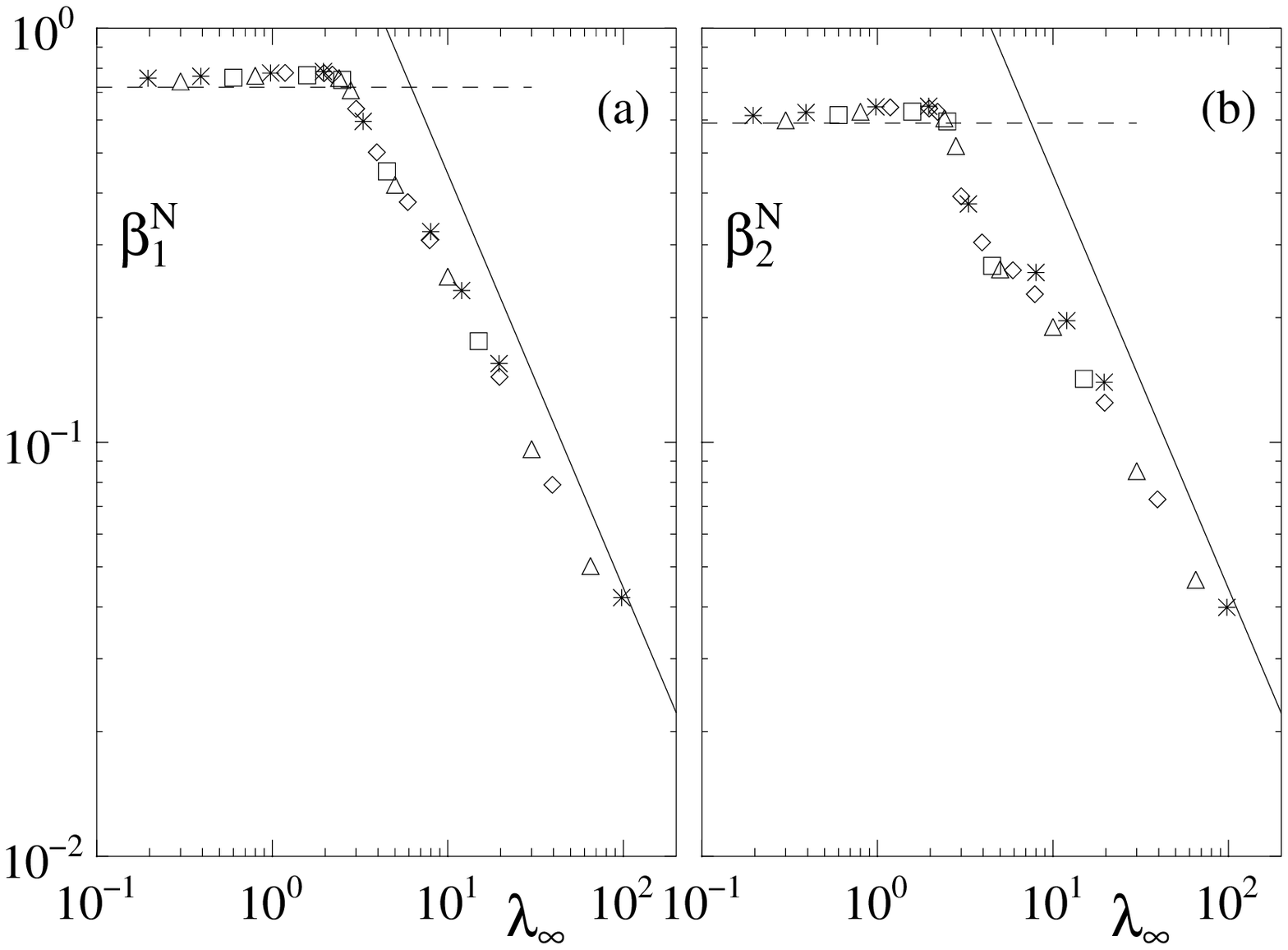,width=8cm}\hfill\\
\end{center}
{\footnotesize {\bf FIG. 3.}
Finite sample scaling of $\beta_q^N$ as a function of $\lambda_\infty$
with $\lambda_N=1$. Different symbols correspond to various sample 
sizes~$N\in[200,1000]$ and disorder strengths~$W\in[0.3,0.6]$; the electric
field was chosen appropriately ($eF\in[5\cdot10^{-4},5\cdot10^{-1}]$).
Dashed lines correspond to Eq.~(\ref{plato}), where the values~$c_q$ were taken from 
a least squares fit of Fig.~4a,b (see below). Full lines represent the scaling law 
derived in Eq.~(\ref{naeherung2}), where the values for~$a_q^0$ where taken from Fig.~2.
{\bf (a)} Scaling of~$\beta^N_1$ ($a_1^0=4.45,\, c_1=2.59$).
{\bf (b)} Scaling of~$\beta^N_2$ ($a_2^0=4.43,\, c_2=1.45$).
}
\end{figure}

Once the electric field is turned on, however, a new length scale~$l_{\rm el}$ 
(with respect to the standard Anderson models) appears.
This can be seen already from the previous paragraph, where on the basis of numerical 
results we were able to conclude that for infinite 1D Anderson models in the presence of 
an electric field the scaling properties of the eigenvectors are characterized by the 
parameter $\lambda_{\infty}$. Since the sample size now enters as a third length scale, 
the second scaling parameter $\lambda_N$ is likely to show up. Thus we expect that the
statistical properties of the eigenstates, and accordingly the $\beta_q^N$'s, 
are going to be determined by the two parameters $\lambda_{\infty}$ and $\lambda_N$
which arise due to the competition between the characteristic lengths $l_{\infty}$ 
and $l_{\rm el}$ of the corresponding infinite sample and the actual size~$N$ of the 
sample. In the rest of the Section we are going to present numerical evidence that 
for finite 1D Anderson models in the presence of an electric field, the statistical 
properties of the eigenstates are characterized by the two scaling parameters 
$\lambda_{\infty}$ and $\lambda_N$. To this end, we will concentrate on the 
localization measures $\beta^N_q$ (see Eq.~(\ref{betaq})), which are the finite size 
counterparts of $\beta_q^\infty$. 

To find the localization lengths $L_q$ for finite samples of size~$N$, we have
used the same approach as in the previous subsection. The average values of $L_q$
were calculated by choosing only the eigenstates which had eigenvalues within
a small energy window $E\in [0.95,1.05]$. Additionally, we performed an 
ensemble averaging over at least 100 realizations of the disorder potential.
For each~$N$ the total averaging thus involved several hundreds up to
several thousands of eigenstates. In all these calculations 
the sample size was varied from $N=200$ up to $1000$.

To test the scaling assumption (\ref{scl}) for finite systems we first analyze
the behaviour of the localization measures $\beta^N_{1,2}$ once $\lambda_N$
is fixed. This is the finite sample counterpart of the scaling analysis presented in the
previous subsection. In Fig.~3 we report our numerical results. Different symbols 
correspond to various sample sizes $N$ and disorder strengths $W$ such that always 
$\lambda_N=1$. The good overlap confirms the scaling dependence $\beta_q^N = 
\beta_q^N(\lambda_{\infty})$ conjectured in Eq.~(\ref{scl}). 

Let us now try to gain some insight in the asymptotic form of the scaling law of
$\beta_q^N (\lambda_{\infty},\lambda_N=const)$. For $\lambda_\infty\ll 1$, the system 
is in the Anderson regime, where $\beta_q^N$ is given by Eq.~(\ref{fse}). The latter
expression does not depend on~$\lambda_\infty$, and thus we can conclude that 
$\beta_q^N$ has to saturate to a constant which is given by 
\begin{equation}
\label{plato}
\beta_q^N(\lambda_\infty\ll 1, \lambda_N)\approx\frac{c_q\lambda_N}{1+c_q\lambda_N}\,\, 
\end{equation}
where $\lambda_N = const.$ The dashed lines in Fig.~3 show the expected saturation 
plateau given by (\ref{plato}) for $\lambda_N =1$. The agreement with the numerical 
data is very good.

In the opposite limit $\lambda_\infty\gg 1$, Stark localization sets the dominant 
length scale~$l_{\rm el}$. Since we can always choose the strength of the electric
field $F$ such that $N,l_\infty\gg l_{\rm el}$, and assuming continuity in the
form of $\beta_q^N$, we can approximate the latter with the help of Eq.~(\ref{fit1}).
For $\lambda_\infty\gg 1$ this formula yields $\beta^\infty_q \approx a^0_q$. Next, by
changing variables and going from $\beta^\infty_q$ to $\beta_q^N$ we get
\begin{equation}
\label{naeherung1}
\beta_q^N=\frac{l_{\rm el}\beta^\infty_q}{N}=\frac{\lambda_Na^0_q}
{\lambda_\infty}\,\, .
\end{equation}
Displaying~$\beta^N_q$ versus~$\lambda_\infty$ in a double logarithmic fashion, 
this yields
\begin{equation}
\label{naeherung2}
\ln(\beta_q^N)\approx \ln(\lambda_N a^0_q)-\ln(\lambda_\infty)\,\, .
\end{equation}
This linear behaviour~(\ref{naeherung2}) is shown by solid lines
in Fig.~3, it describes approximately the numerical data. Deviations are due to 
the fact that the 
approximation via the scaling law of Eq.~(\ref{fit1}) actually requires not only
$l_{\rm el}\ll N$ but also $l_\infty\ll N$, which is not fulfilled in our case. 
Nevertheless it gives a reasonable estimate.

We now turn to the case, where~$\lambda_\infty$ is fixed and~$\lambda_N$ varies.
Our numerical results, for $\lambda_{\infty} = 0.01,1,20$, corresponding to the 
Anderson, intermediate and Stark regime, respectively, are shown in Fig.~4 where
now we refer to the new variable
\begin{equation}
Y_q = \frac{\beta_q}{1-\beta_q}\, .
\end{equation}
The points corresponding to the same $\lambda_{\infty}$ (but different $l_{\infty}$ 
and $l_{\rm el}$) fall onto the same smooth curve with a good accuracy, confirming 
the scaling hypothesis (\ref{scl}). From Fig.~4 one can see that the
behaviour of $Y_q$ is different in the two regimes $\lambda_{\infty}\gg1$
($\lambda_{\infty}\ll 1$) where localization is due to the Stark (Anderson)
mechanism. As a matter of fact, as we are increasing~$\lambda_{\infty}$
two asymptotic regimes start to build up which have the same slope
and differ only by a constant shift.

To understand, the behaviour of $Y_q(\lambda_N)$ as a function of $\lambda_{\infty}$,
we first try to illuminate the limiting cases. Let us start with the limit
$\lambda_\infty\ll 1$. This condition, defines the Anderson regime, where 
Eq.~(\ref{fse}) holds and thus a behaviour 
\begin{equation}
\label{Ylo}
\ln(Y_q)=\ln(\lambda_N)+\ln c_q
\end{equation}
in terms of the new variable $Y_q$
is expected over the whole range of~$\lambda_N$. This expectation is shown in Fig.~4
by solid lines. 
Since~$\lambda_\infty$ is small but still finite we can estimate the~$c_q$ also 
from Eq.~(\ref{fit1}). The limit~$\lambda_N\ll\lambda_\infty\ll 1$ corresponds to 
the Anderson regime of a nearly infinite sample. Expanding Eq.~(\ref{fit1}) to first 
order yields~$\beta_q^\infty\approx a^0_qa^1_q\lambda_\infty$. Substituting this 
expression into Eq.~(\ref{naeherung1}) we end up with the following 
term for $\beta_q^N$
\begin{equation}
\label{matth}
\beta_q^N\approx a^0_qa^1_q\lambda_N\,\, .
\end{equation}
Inserting (\ref{matth}) into the definition of~$Y_q$ and assuming $\lambda_N\ll 1$ 
we get 
\begin{equation}
\label{gerade}
\ln(Y_q)\approx\ln(\lambda_N)+\ln(a^0_qa^1_q)\quad, 
\end{equation}
which implies~$c_q\approx a^0_qa^1_q$. A comparison of $c_q=2.59\,(1.55)$ and
$a^0_qa^1_q=2.45\,(1.64)$ shows a very good agreement. Thus, we conclude that 
our scaling function (\ref{fit1}) is consistent with Eq.~(\ref{Ylo}) as it should
be in the above limit.

\begin{figure}
\begin{center}
\epsfig{figure=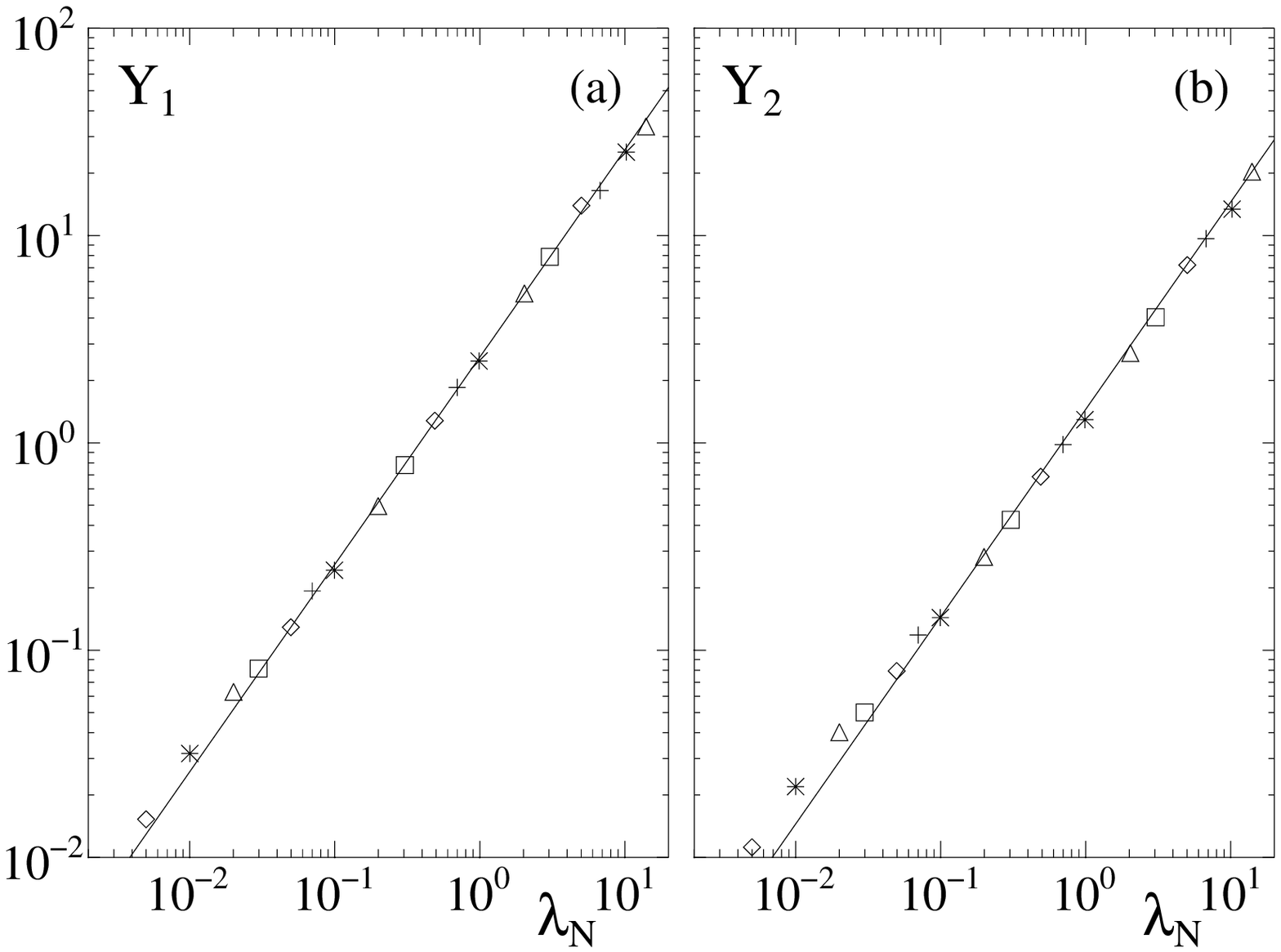,width=8cm,angle=360}\hfill\\
\epsfig{figure=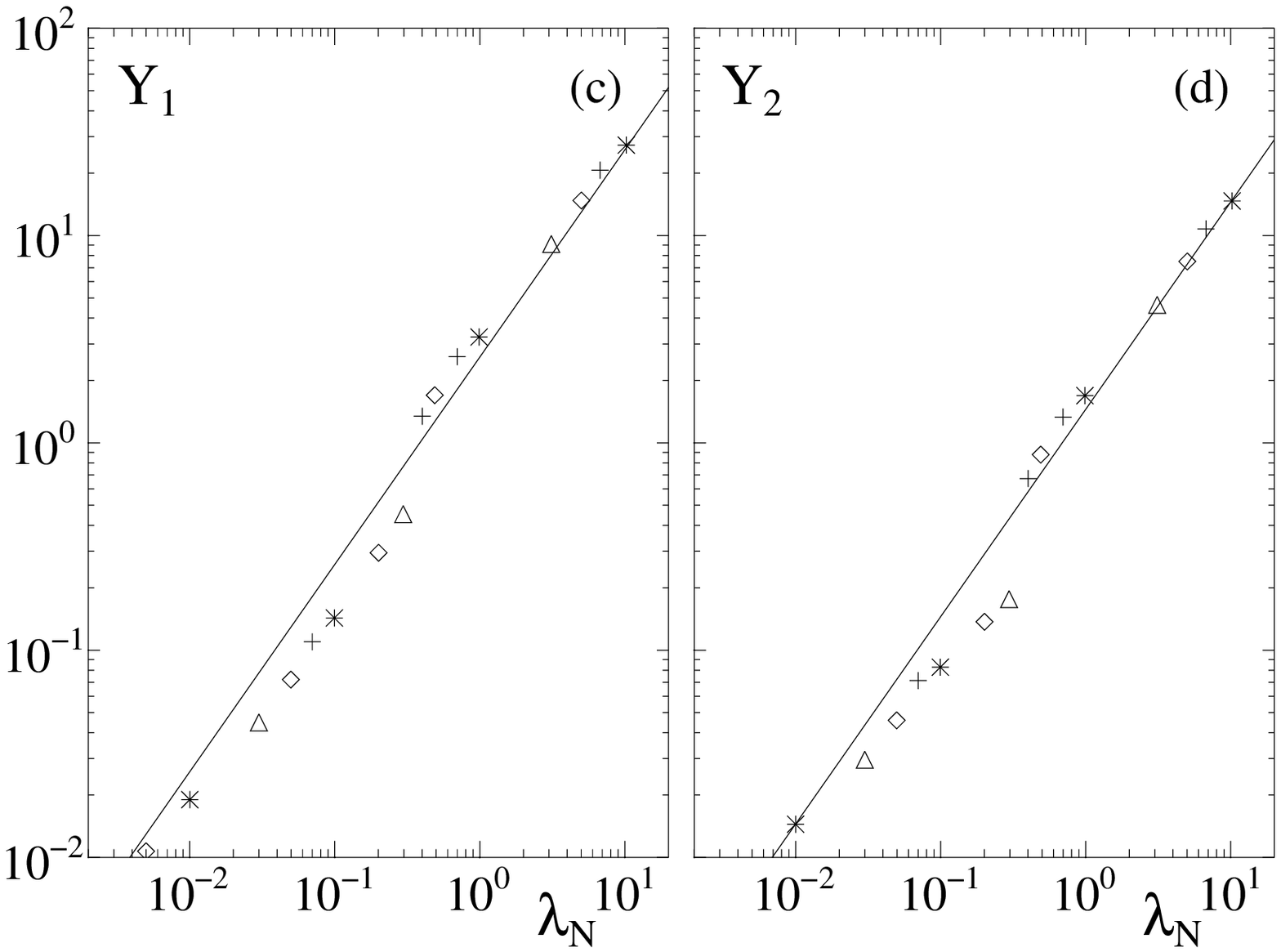 ,width=8cm,angle=360}\hfill\\
\epsfig{figure=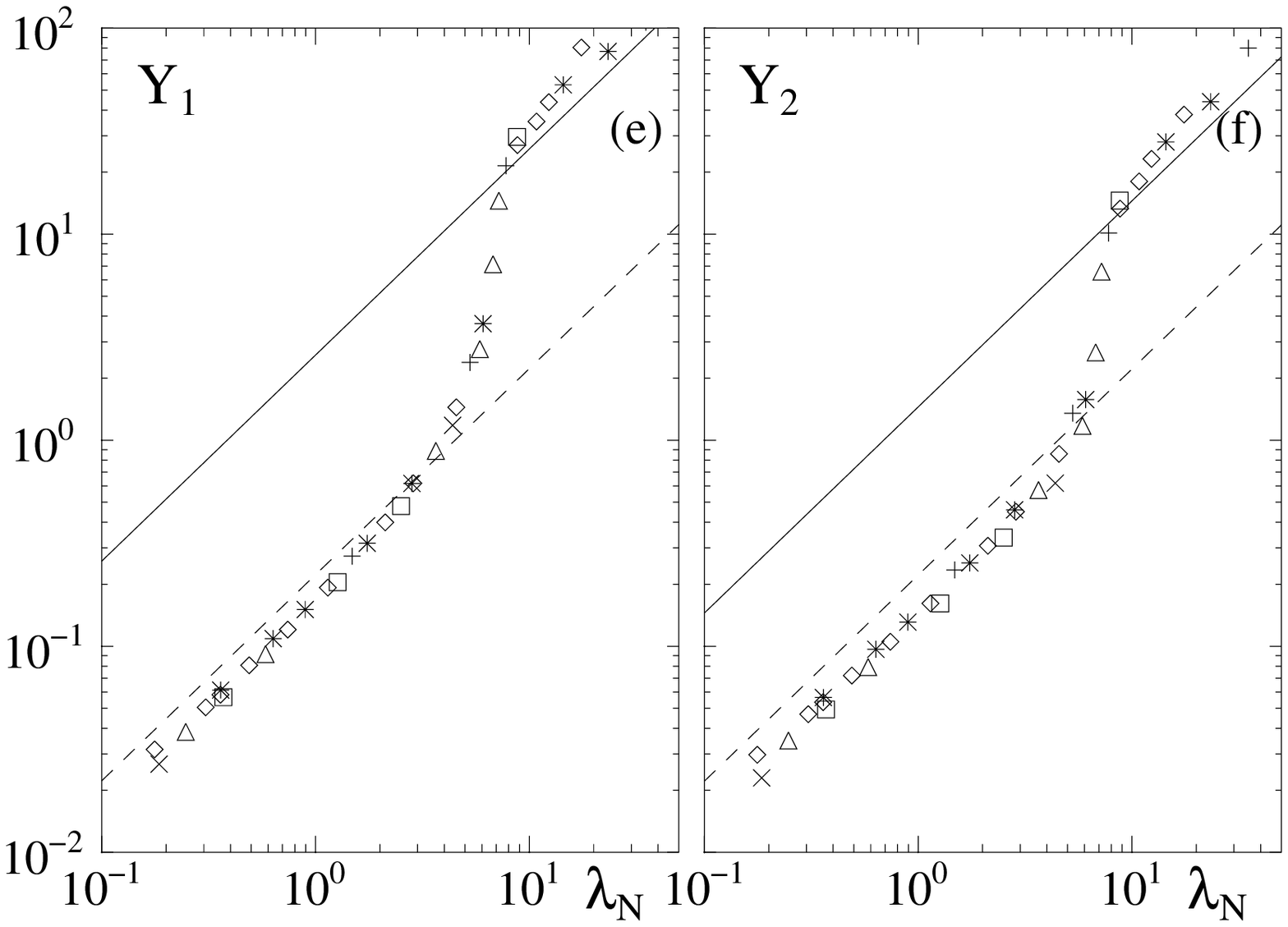,width=8cm,angle=360}\hfill\\
\end{center}
{\footnotesize {\bf FIG. 4.} Scaling of $Y_q = \frac{\beta_q}{1-\beta_q}$ 
in the finite sample upon variation of $\lambda_N$ for $\lambda_{\infty} =0.01,1,20$. 
Different symbols correspond to various disorder strengths~$W\in[0.1,6]$ and sample sizes
$N\in[200,1000]$. Full and dashed lines 
correspond to the limiting cases given by Eq.~(\ref{fse}) (with $c_1=2.59,\,c_2=1.45$) and
Eq.~(\ref{gerade2}) (with $a^0_1=4.45,\,a^0_2=4.35$), respectively.
{\bf (a,b)} Scaling of~$Y_1,Y_2$ for~$\lambda_\infty=0.01$ (Anderson regime)
{\bf (c,d)} Scaling of~$Y_1,Y_2$ for~$\lambda_\infty=1$ (intermediate regime)
{\bf (e,f)} Scaling of~$Y_1,Y_2$ for~$\lambda_\infty=20$ (Stark regime)
}
\end{figure}

In the opposite limit of~$\lambda_\infty\gg1$, we have to distinguish between the
following two cases. When ~$\lambda_N\gg\lambda_\infty$, the sample size~$N$ sets the 
smallest length scale. In this limit the eigenstates extend over the whole sample.
Then by continuity we expect that the behaviour of $Y_q(\lambda_N)$ for $\lambda_N 
\gg 1$ will be given by Eq.~(\ref{Ylo}). Our numerical data (see Figs.~4e,f)
support this hypothesis. The second case, in which $\lambda_N\ll\lambda_\infty$ holds,
can be viewed as the extreme Stark regime of an infinite sample. In that limit one 
obtains Eq.~(\ref{naeherung2}) again, which yields 
\begin{equation}
\label{gerade2}
\ln(Y_q)\approx\ln(\lambda_N)+\ln\left (\frac{a^0_q}{\lambda_\infty} \right )\quad .
\end{equation}
The asymptotic behaviour ~(\ref{gerade2}) is reported in Figs.~4e,f with dashed lines
and agrees quite well with our numerical data.

From the above analysis we conclude that at
\begin{equation}
\label{asym}
\lambda_{\infty} \simeq \lambda_N
\end{equation}
two asymptotic regimes are created due to the interplay between Anderson and Stark
localization mechanisms. Although these estimations are made only on a very rough level, 
they describe our numerical findings rather well.

%----------------------------------------------------------------------------
\section{Scaling of the distribution of eigenvectors}
%----------------------------------------------------------------------------

As a complement to the above analysis, we show in this section, that the distributions 
of squared components of eigenvectors are parametrized in the same fashion by
$\lambda_N$ and $\lambda_\infty$. Again we restrict ourselves to a definite energy 
window $E\in [0.95,1.05]$, where all eigenvectors corresponding to eigenvalues in that 
window were computed for several realizations of disorder. The total number of 
eigenstates were in all cases more than $1000$. 

Before examining the scaling properties of the distribution let us distinguish
between the various cases that appear due to the competition between the three
characteristic length scales $N,l_{\rm el},l_\infty$ (see previous section).
For simplicity we define these regimes only by their limiting cases, which
read as follows:
\begin{enumerate}
\item $l_\infty\! <\! l_{\rm el}\! <\! N\,\,$ \& $\,\,l_{\rm el}\! <\! l_\infty\!
<\! N\,\,$ (infinite sample)
\item $l_\infty\! <\! N\! <\! l_{\rm el}\,\,$ \& $\,\,N\! <\! l_\infty\! <\!
l_{\rm el}\,\,$ (Anderson regime)
\item $N\! <\! l_{\rm el}\! <\! l_\infty\,\,$ \& $\,\,l_{\rm el}\! <\! N\! <\!
l_\infty\,\,$ (Stark regime)
\end{enumerate}

The first pair in this list corresponds to scaling behaviour of the infinite
sample, since the sample size is always larger than the other two competing
lengths. For this case and~$q=1,\,2$ we have shown already in Section~IV.A that 
$\beta_q^{\infty}$ follows a single parameter scaling with respect to~$\lambda_\infty$. 
We will show here that also the whole distribution of eigenvector components is 
parametrized according to the same scaling parameter. The first question which 
arises for the infinite sample, is the proper normalization of the squared entries 
of the eigenstates~$w_n=|\varphi_n|^2$. A normalization with respect to the number 
of sites~$N$ does not seem appropriate since we are interested in the limit $N\rightarrow 
\infty$. Since the~$w_n$ have to scale with some length, however, following our previous 
strategy (see Eq.~(\ref{betaq})), we introduce the variable
\begin{equation}
\label{dis1}
r=\ln (w_nl_{\rm el})
\end{equation}
and investigate the distribution~$p(r)$. For our calculations we consider
matrices of size $N=10^4$, while we $l_{\rm el},l_\infty\ll N$ and varied  
$\lambda_\infty$. For each~$\lambda_\infty$ we considered two different disorder 
strengths~($l_\infty\approx 6, 10$) and adjusted the dc field strength appropriately. 
The results for $\lambda_\infty=10,1,0.1$ are shown in Fig.~5a-c in a 
semilogarithmic plot. The assumed scaling of $p(r)$ with $\lambda_{\infty}$
is clearly visible. 

\begin{figure}[t]
\begin{center}
\epsfig{file=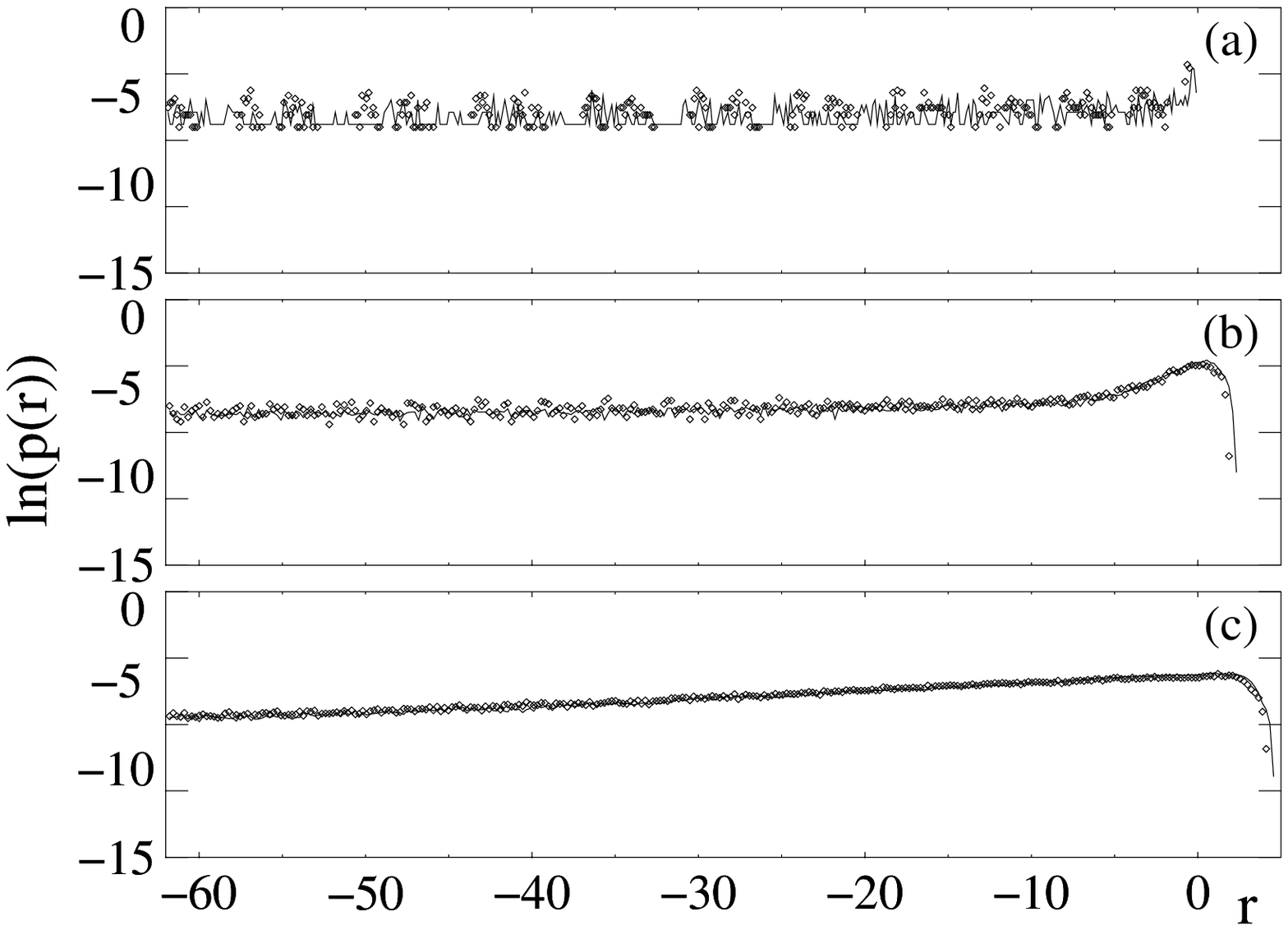,width=8cm}\hfill\\
\end{center}
{\footnotesize {\bf FIG. 5.}
Scaling of the entire distribution of squared eigenvector components $p(r)$
with~$\lambda_{\infty}$ in the case of nearly infinite samples ($N=10^4$).
Two different pairs of~$l_\infty,\,l_{\rm el}\ll N$ are denoted by full lines 
($W=3.49, eF=1.666,0.166,0.016$) and symbols ($W=2.62, eF=1,0.1,0.01$), 
while keeping~$\lambda_\infty$ fixed.
{\bf (a)} $\lambda_\infty=10$
{\bf (b)} $\lambda_\infty=1$
{\bf (c)} $\lambda_\infty=0.1$
 }
\end{figure}

The second and third pair of conditions always involve the sample size~$N$.
Therefore scaling according to~$\lambda_\infty$ {\it and}~$\lambda_N$ has to
be taken into account. For these cases renormalization with respect to the
sample size is meaningful; hence we define the rescaled squared entries of
eigenstates as
\begin{equation}
\label{dis2}
r=\ln (w_nN).
\end{equation}
Before turning to the analysis of our numerical data, let us first qualitatively
analyze the form of $p(r)$. In the limit $N\ll l_\infty, l_{el}$ the
system does not feel any localization due to disorder or electric field.
Therefore all eigenstates are given approximately by Eq.~(\ref{EF}).
The distribution $p(r)$ is then given by 
\begin{equation}
\label{dis3}
p(r)=\frac{e^r}{\pi\sqrt{e^r(2-e^r)}}\quad .
\end{equation}
Plotting $\ln(p(r))$ versus~$r$ for~$r\ll 0$ yields a curve with slope~$1/2$, 
as can be verified by expanding Eq.~(\ref{dis3}). Moreover Eq.~(\ref{dis3}) 
shows a sharp peak around~$r=0$.

In the case where $l_\infty\ll N, l_{\rm el}$ the disorder sets the relevant
length scale and the system resembles an infinite Anderson model with
exponentially localized eigenstates~$w_n=\exp(-\vert n-n_0\vert /l_\infty)$. 
For small~$r$ this particular form of eigenstates leads to \cite{M93}
\begin{equation}
\label{dis4}
p(r)=l_\infty/N.
\end{equation}
In a semilogarithmic representation this results in a nearly horizontal curve
of height~$\ln(l_\infty/N)$ for~$r\ll 0$, which drops rapidly for some~$r>0$. 

\begin{figure}[t]
\begin{center}
\epsfig{file=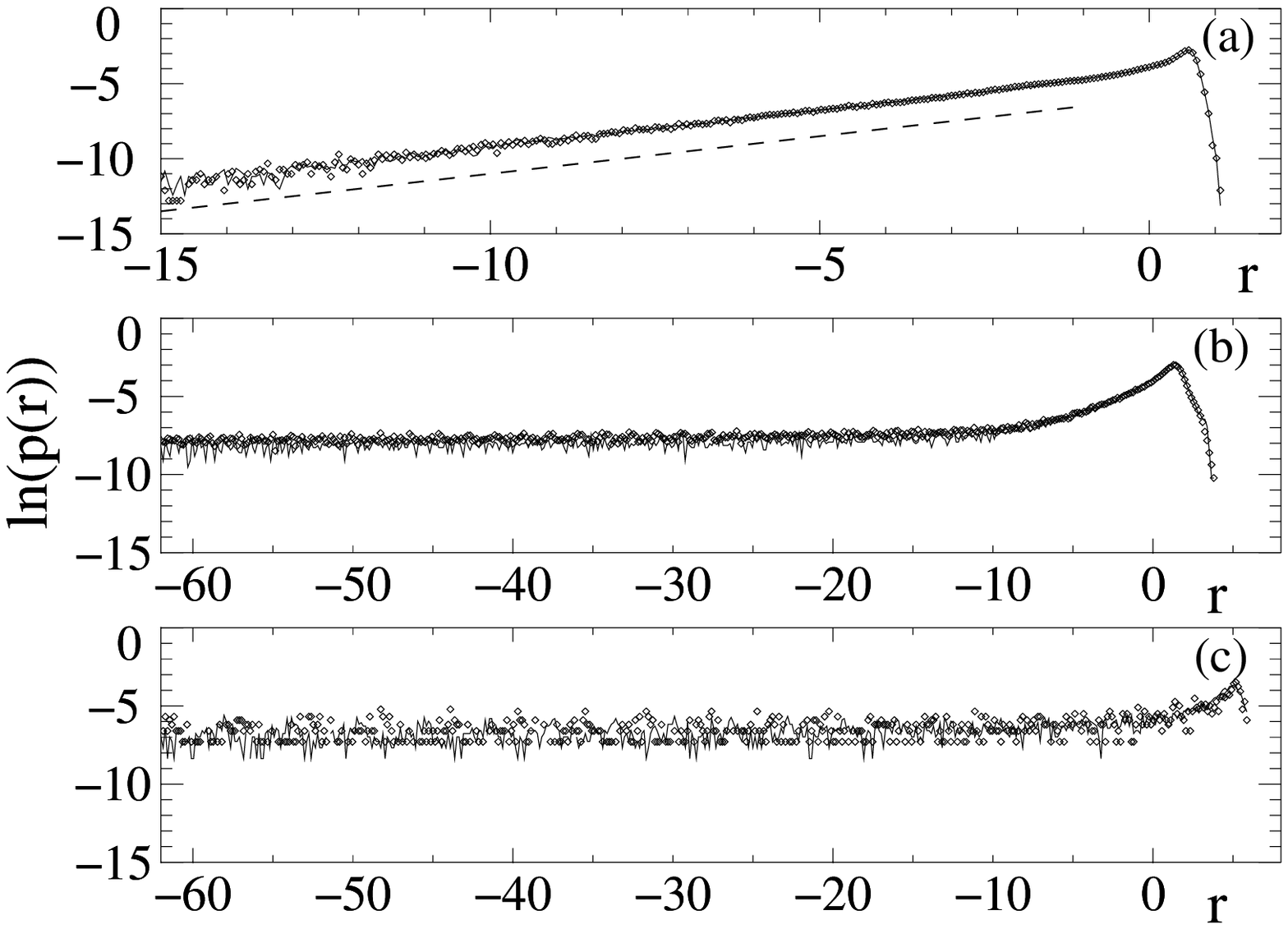,width=8cm}\hfill\\
\end{center}
{\footnotesize {\bf FIG. 6.}
Scaling of the entire distribution of squared eigenvector components $p(r)$ 
in the Stark regime~($\lambda_\infty=20$) in the case of finite samples
($N\in[230,1200],W\in[0.01,2],eF\in[10^{-3},1]$). Full lines and symbols
denote different sample sizes (and thus different strengths of disorder), 
while~$\lambda_N$ is kept fixed to a chosen value.
{\bf (a)} $\lambda_N=100$ (the dashed line has slope~$1/2$)
{\bf (b)} $\lambda_N=2$
{\bf (c)} $\lambda_N=0.05$
}
\end{figure}

For the second pair of conditions, i.e. the Anderson regime, the scaling 
properties of the distribution were already analyzed in~\cite{M93}. A good 
agreement with the limiting Eqs.~(\ref{dis3}),(\ref{dis4}) was found. 

The new and more interesting case is the pair of conditions with label~3,
where the competition between~$N$ and~$l_{\rm el}$ is dominant. In this case
$\lambda_{\infty}\gg 1$ and thus the localization mechanism is due to the Stark 
effect. The resulting distribution for some representative values of 
$\lambda_{\infty}$ and $\lambda_N$ are shown in Fig.~6. One can clearly see
that distributions corresponding to different sample sizes $N$ and disorder
strengths $W$, but having the same $\lambda_N$ and $\lambda_{\infty}$ coincide.
Moreover, as we move from higher to smaller values of~$\lambda_N$
the shape of the distributions changes drastically. In the case $\lambda_N\gg1$
(corresponding to $N\ll l_{\rm el}$) the eigenstates can be considered as 
extended with respect to the sample size, and thus we obtain again 
Eq.~(\ref{dis3}) (see Fig.~6a). The peak of the distribution broadens and 
moves to the right upon an increase of~$\lambda_N$ as can be seen from 
Fig.~6b. At the same time, for negative values of $r$ the distribution 
possesses long tails. This becomes more and more apparent as we are moving 
towards the Stark regime (Fig.~6c). In the strong field limit the eigenstates 
are essentially localized at one site i.e. $w_n \simeq \delta_{nm}$. In this 
case, one has $p\left(r\simeq ln(N)\right) \sim 1/N$, while long tails appear
due to factorial localization.

We conclude this Section by noticing that the scaling of the distributions 
of squared eigenvector components with $\lambda_{\infty}$ and $\lambda_N$ implies
scaling of the localization parameters~(\ref{scl}) 
for arbitrary $q$.

%------------------------------------------------------
\section{\bf Summary}
%------------------------------------------------------

We have studied a 1D tight-binding model in the presence of a constant 
electric field. For such a model we can distinguish between two regimes 
where localization is due to totally different mechanisms. The first regime 
is the Anderson regime, which is defined through the condition 
$\lambda_\infty\ll 1$. Here the localization due to disorder is the dominant 
mechanism that controls the statistical properties of the eigenstates. In 
the opposite limit, $\lambda_\infty\gg 1$, the localization is due to the 
presence of the electric field. This is the Stark regime. Our numerical 
study deals with the scaling properties of the eigenstates both for infinite 
and finite samples. 
This study was motivated by the remarkable scaling law that has been found 
for quasi-1D models with electric field~\cite{FM91,CCGI93,FGIM93}.  
Our results indicate that in both infinite and finite samples with disorder 
and electric field the eigenstates have generic properties, regardless of 
the dimensionality of the system, provided that an appropriate renormalization 
(with respect to the corresponding "extended'' states) is done. Thus we show 
that the similarity between 1D and quasi-1D eigenstates should persist also
for systems with electric field.

We found that for infinite systems the statistical properties of the
eigenstates are characterized by the single parameter $\lambda_{\infty}$. 
This conclusion was based on an extensive numerical analysis of both the 
localization measures (\ref{betaq}) and of the whole distribution of squared
eigenvector components. Moreover for $\beta_{q=1,2}^{\infty}$ we have found 
a simple scaling law (\ref{fit1}) that describes quite nicely our numerical 
data. This expression can be used in order to find the strength of the applied 
dc electric field once~$l_{\infty}$ is known for the field free model.

Moreover, we have performed a finite length scaling analysis. Our numerical
analysis showed that for finite systems the statistical properties of the
eigenstates are characterized by two parameters, namely $\lambda_{\infty}$,
as in the case of infinite systems, and $\lambda_N$. The latter parameter involves
the actual size of the sample which enters in the scaling analysis as a
third length scale. We found that the localization measures $\beta_{q=1,2}^N$ 
show a totally different asymptotic behaviour in $\lambda_N\rightarrow 0,\infty$
as we are increasing the parameter $\lambda_{\infty}$. Based on some analytical 
arguments we estimated that this occurs approximately at $\lambda_N \simeq 
\lambda_{\infty}$. This can be used as a criterion to identify which
localization mechanism (Anderson or Stark localization) is responsible for
the structure of eigenstates. It will be interesting to investigate if
the same asymptotic behaviour also holds for higher moments~$q$. Finally,
we studied the whole distribution of squared eigenvector components and 
showed that it also follows the same scaling behaviour with respect to
the two scaling parameters $\lambda_{\infty}$ and $\lambda_N$.

The main result of our work is the fact that scaling properties of
eigenstates of infinite systems are described by one parameter scaling 
$\lambda_{\infty}$, whereas for finite systems an additional parameter $\lambda_N$ 
is also needed. 
In particular, both localization lengths, the entropy localization length as well as
the one defined via the inverse participation ratio, follow the universal scaling law 
of Eq.~(\ref{scl}) after appropriate normalization.
This is in contrast to the
standard Anderson models without electric field, where only
one parameter~($\lambda_N$) is needed to describe the scaling properties of eigenstates.

%-----------------------------------------------------------------------------------

\end{multicols}
\end{document}